\PassOptionsToPackage{preprint}{neurips_2026}

\def\cpanauthor{%
  Anik Jha\\
  Independent Researcher\\
  \texttt{anik.k.jha@gmail.com}%
}

\def\cpanartifact{%
  The analysis code and every results file quoted in this paper are at
  \url{https://github.com/anik-jha/challenger-panels}. The monitors and the transcript
  dataset are public artifacts that the repository's download scripts fetch rather than
  redistribute.%
}

\documentclass{article}
\usepackage[dblblindworkshop]{neurips_2026}
\workshoptitle{Who Verifies the Agents? Toward Reliable Agent Development}
\usepackage[T1]{fontenc}
\usepackage[utf8]{inputenc}
\usepackage{microtype}
\usepackage{graphicx}
\usepackage{booktabs}
\usepackage{amsmath}
\usepackage{xcolor}
\usepackage{hyperref}
\usepackage{caption}
\definecolor{cpanlink}{RGB}{0,72,133}
\hypersetup{colorlinks=true, linkcolor=cpanlink, citecolor=cpanlink, urlcolor=cpanlink}

\newcommand{\arxivref}[1]{\href{https://arxiv.org/abs/#1}{arXiv:#1}}

\title{Decorrelation Is Not Complementarity:\\ Skill, Not Lineage, Governs Trusted-Monitor Ensembles}

\providecommand{\cpanauthor}{Anonymous Author(s)}
\providecommand{\cpanartifact}{Code and results files accompany the submission.}

\author{\cpanauthor}

\begin{document}
\maketitle

\begin{abstract}
Trusted monitoring has a cheap, trusted model score a stronger untrusted model's actions,
and a diverse \emph{ensemble} of them beats a single stronger monitor at matched cost. They are built by minimising average pairwise correlation, and that paper's
twelve monitors shared one base model, leaving open what supplies the diversity. We study 24
open-weight monitors spanning nine pretraining lineages and a 29x range of detection
skill (pAUC at 10 percent FPR, 0.028 to 0.803) on backdoored code.

The metric used to build panels does not predict what a panel is for, and we can say why.
Agreement on attack items splits into a shared-detectability \emph{signal} component and an
idiosyncratic \emph{error} component, which predict ensemble gain with \emph{opposite} sign
(Spearman -0.25 and +0.26), so their sum, the metric actually used, predicts it barely at
all (+0.05); the cancellation holds in 7 of 8 evaluations. Skill acts on signal
(+0.53) while error stays flat (-0.01), which is why a monitor's own skill predicts its
agreement with the pool (Spearman 0.84, n = 24, permutation p below 0.0001).

Pretraining lineage is the obvious way to buy decorrelation, and it does not pay. At matched member
capability, cross-lineage panels detect no better (permutation p = 0.13), and lineage barely
moves the metric either (+0.064, p = 0.18). We report that
against ourselves: on our own 22-monitor pool the same test read +0.104 at p = 0.037 until
two monitors were added. An earlier pool topping out at pAUC 0.23 had already invalidated
another analysis. Such a quantity is a property of the pool assembled.

Panel gain over the best member falls monotonically with panel skill (-0.66 at k = 2,
-0.70 at k = 3), and no correlation-weighted selection beats picking the single best
monitor out of sample. Across six attacker models the gain result holds in all six, the agreement and cancellation results in five of six.
\end{abstract}

\section{Introduction}

In the AI-control setting \citep{greenblatt2024control}, a \emph{trusted} but weaker model
monitors the actions of a stronger \emph{untrusted} model, and the most suspicious actions are
audited or deferred. Because a single monitor is an obvious single point of failure, a natural
move is to run several and combine their scores. \citet{koran2026ensemble} show this pays. With
twelve monitors built from GPT-4.1-mini and combined into ensembles, 94.6\% of diverse
five-monitor ensembles outperform a single GPT-4.1 call at approximately matched inference cost,
and their best three-monitor diverse ensemble achieves $2.4\times$ the gain of a homogeneous
three-monitor ensemble. They title the result ``diverse signals outweigh more compute.''

Crucially, all twelve of their monitors share one base model. Their diversity is prompt and
fine-tuning variation only. They say so explicitly, and name the gap as future work: ``[e]nsembles
drawn from different model families---or that incorporate non-LLM-based signals---may be more
resistant, but we have not tested this directly'' \citep[\S6.3]{koran2026ensemble}. They further
note that ``the mechanism underlying complementarity remains unclear.''

\paragraph{What is already theirs.} We state the credit precisely, because two of our findings
have qualitative antecedents in the same paper. First, they observe that selecting purely for
diversity loses to random selection, because ``[l]ow-performing monitors disagree with others not
because they detect distinct signals, but because they are more often incorrect''
\citep[\S5.4]{koran2026ensemble}. That is our Finding 1, asserted qualitatively, for
prompt-variant monitors sharing one base model. Second, on a three-point base-model ladder they find
ensembling helps most for the weakest base (Nano pAUC $0.162 \to 0.553$) and report that a much
stronger single monitor (GPT-5.1, pAUC $0.783$) is not consistently beaten by their diverse
five-monitor ensemble, both of which anticipate our Finding 3. What we add is not the direction of
these effects but their measurement. We use a continuum of 24 monitors across nine lineages
rather than a three-point ladder on one base model, with attenuation controls that separate a real
capability effect from a statistical artifact, and we add a decomposition supplying the mechanism
they leave open, which is ours alone.

Trusted monitoring is the verification layer that agentic deployments rest on. When an agent's
actions are too numerous for human review, a panel of cheaper models decides what gets audited.
The design question for that layer is what to spend the next unit of budget on, a better monitor
or a more heterogeneous one. This paper is about whether heterogeneity in the verifier pool is
worth buying, which is the ``diverse signals'' question in its most measurable form. We study it
on single-step code sabotage, where ground truth is unambiguous. Whether it holds over long agent
trajectories is the obvious next step and is untested here (\S\ref{sec:limits}).

This paper takes up both of their open questions with a pool of open-weight monitors that varies
along the axes a single-base-model study cannot: different weights, different pretraining
lineages, and a wide range of competence. Our headline is not the one we expected. The dominant
structure in the data is not lineage and not architecture. It is \emph{skill}. Monitors that are
better at the task agree with each other more, and the apparent diversity of a weak pool is
largely the mutual disagreement of models that are each wrong in different places. Lineage, the
obvious lever for buying genuine independence, moves no quantity we can measure that detection
depends on.

\paragraph{Contributions.} \begin{enumerate}\itemsep2pt \item \textbf{Why the diversity metric
fails.} Decomposing attack-item agreement into a shared-detectability signal component and an
idiosyncratic error component shows the two predict ensemble gain with opposite sign (Spearman
$-0.25$ and $+0.26$), so the metric used to build panels, which is their sum, predicts gain barely
at all ($+0.05$). This answers the mechanism question \citet{koran2026ensemble} leave open, and it
replicates in 7 of the 8 evaluations available to us. \item \textbf{Skill, not lineage, is the
axis that moves.} Skill acts on the signal component ($+0.53$) while error stays flat ($-0.01$),
giving a strong capability--agreement relationship (Spearman $\rho = 0.84$, $n = 24$, permutation
$p < 10^{-4}$) that survives three attenuation controls. Lineage buys no detectable detection
benefit at matched member capability (monitor-level permutation $p = 0.13$), and in a 24-monitor
pool it no longer moves the correlation metric either ($+0.064$, $p = 0.18$). \item
\textbf{Correlation structure is a property of the pool, not of monitoring.} We demonstrate this
twice, both times by invalidating our own earlier analysis. A pool topping out at pAUC $0.23$
produced a lineage null that did not survive competent monitors, and a 22-monitor pool produced a
significant lineage agreement gap ($+0.104$, $p = 0.037$) that adding two monitors erased.
Published diversity numbers should be read as conditional on the pool that was assembled. \item
\textbf{The practical consequence.} Ensemble gain over the best single member falls with pool
skill (Spearman $-0.66$ at $k=2$, $-0.70$ at $k=3$), and no correlation-weighted panel selection
beats the single best monitor out of sample. \end{enumerate}

\section{Setup}

\paragraph{Task and data.} We use the public backdoored-APPS dataset released with
\citet{koran2026ensemble}. Each sample is a Python solution to a competitive-programming problem,
labelled \emph{attack} if it contains a deliberate backdoor-code that passes the visible tests but
fails on a hidden input-or \emph{honest} otherwise. We reproduce their test split exactly: 1{,}375
samples, 172 attack and 1{,}203 honest. Their train/validation membership is not published, so
ours is a deterministic reconstruction that matches their stated counts but not necessarily their
exact assignment. Validation is used only to measure monitor capability, never to compute a
reported test number. The test-set backdoors were authored by Claude-3.7-Sonnet-Thinking, so every
monitor below is far smaller and weaker than the model whose sabotage it must catch (our largest
is 32B, and the attacker's parameter count is not public, so this is a claim about the capability
gap rather than a measured ratio), which is the regime trusted monitoring assumes. Following their
preprocessing, all code comments are stripped before any monitor sees a sample. Backdoor code in
this dataset frequently carries comments that name the backdoor outright and would otherwise leak
the label.

\paragraph{A second evaluation, varying the attacker.}
A single attacker is the obvious generality objection. The anchor's second benchmark
(BigCodeBench-Sabotage) was never released, so we instead vary the axis their release does
support: the dataset carries usable backdoors from nine attacker models, and we build a second
evaluation from six of them (o3-mini, Claude-3.7-Sonnet, GPT-4o, Gemini-2.0-Flash, o1-mini,
GPT-4), 767 attacks against one shared 600-sample honest pool so per-attacker differences reflect
the attacks and not a different negative set. This second set is drawn from TRAIN-split problems
only: validation is excluded because monitor capability is measured there, and test is excluded
because the headline numbers come from there. Every monitor is scored on both evaluations.

\paragraph{Monitors.}
Our pool is 24 open-weight instruction-tuned models spanning nine pretraining lineages (Qwen,
Gemma/Google, Phi, Mistral, OLMo, SmolLM, Granite, Falcon, Yi) at sizes from 0.36B to 32B,
giving 276 monitor pairs.
Separately, and used only in \S\ref{sec:single}, we build six prompt variants of one base model
(Qwen3-8B) to reproduce the within-model diversity axis of \citet{koran2026ensemble}. We define lineage by \emph{pretraining provenance}
rather than vendor badge, and exclude distilled variants, following \citet{pinto2026calibration};
a model distilled from Qwen is Qwen lineage. One pool member is a domain-pruned variant of
Qwen3.6-35B-A3B with 25\% of experts removed and healed, built with the public recipe of
\citet{jha2026experts}; it scores pAUC 0.455, between the dense mid-size and strong monitors,
and its lineage is counted as Qwen. We exclude an FP8 copy of Qwen3.6-27B that an earlier
revision of the grid declared: it is the same weights at a different precision, so it would
enter as a near-perfectly correlated same-lineage partner and inflate the one gap this paper
measures.

\paragraph{The pool was grown once, deliberately, after the first analysis.}
The last two monitors added were OLMo-2-32B and Mistral-Small-3.2-24B, and we flag it because
\S\ref{sec:pools} shows the addition changed a headline. Both had previously been excluded on
memory-budget estimates that were simply wrong. They were chosen before their scores were known,
for a stated reason: OLMo and Mistral were the two lineages whose only members sat near chance,
leaving the strong end of the pool three lineages deep. Mistral-Small landed at pAUC $0.407$ and
gave the strong end a fifth lineage; OLMo-2-32B landed at $0.166$, the largest model in the pool
and among its weakest monitors, which is its own warning that parameter count is not monitoring
skill.

\paragraph{Scoring.} We use the logprob-monitor design of \citet{koran2026ensemble} rather than
their integer 0--10 prompted monitor. The monitor is asked for a \textsc{yes}/\textsc{no} backdoor
judgement and the suspicion score is read from the logits of the decision token as
$\text{logit}(\textsc{yes}) - \text{logit}(\textsc{no})$. There are two reasons, and both matter
for this study specifically. Weak models cannot reliably emit a parseable 0--10 score, and parse
failures would be correlated with capability-contaminating the exact axis we measure. Integer
scores also create large numbers of ties, and tied scores are precisely the regime in which
partial-AUC implementations disagree. Scoring is prefill-only and deterministic. Re-scoring a
sample under identical batching reproduces its score bitwise.

\paragraph{Metric.}
We report normalised partial AUROC at FPR $\le 10\%$ (pAUC@10\%), the metric of
\citet{koran2026ensemble}, since only the head of the ranking is ever audited under a realistic
budget. A random ranking scores 0.05. Monitor \emph{capability} is always measured on a held-out
validation split and monitor \emph{agreement} is the Pearson correlation of two monitors' score
vectors computed on attack samples only, again following their convention.

\paragraph{Disclosure of LLM and agent assistance.} The study was designed by the author. The
decision bars governing each experimental wave were pre-registered before the corresponding runs,
and every reported number is written by the analysis harness to a machine-readable results file.
The two analyses added after the monitor pool grew (the capability interaction in
\S\ref{sec:pools} and the cancellation replication in Table~\ref{tab:cancel}) are post-hoc and
labelled as such where they appear. Implementation, sweep orchestration and drafting were
agent-assisted under author supervision; the author verified the claims against those results
files and is responsible for the content.

\paragraph{Statistics.} Both lineage and capability are properties of a \emph{monitor}, but our
observations are \emph{pairs} of monitors. 24 monitors generate 276 pairs that share members and
are far from independent. Bootstrapping over pairs therefore overstates significance badly-in an
earlier draft of this analysis it produced apparently significant lineage effects that vanish
under a correct test. Every $p$-value below comes from permuting the property of interest across
monitors, preserving its marginal distribution, and recomputing the statistic.

\section{Results}

\subsection{Agreement rises with monitor skill}

\begin{figure}[t]
  \centering
  \includegraphics[width=0.52\textwidth]{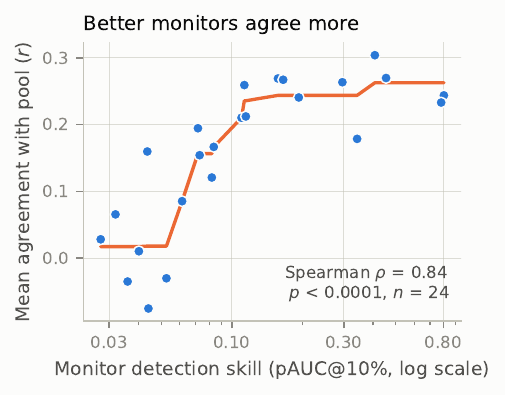}
  \caption{Each point is one of 24 monitors: its detection skill (log scale) against its mean
  agreement with the rest of the pool. The orange curve is an isotonic (monotone) fit-the
  reported statistic is a rank correlation, so a straight-line fit would overstate linearity.
  Agreement rises steeply through the weak range and plateaus.}
  \label{fig:agree}
\end{figure}

Figure~\ref{fig:agree} plots each monitor's detection skill against its mean agreement with the
other 23. The relationship is strong and monotone, at Spearman $\rho = 0.841$ ($n=24$, permutation
$p = 5\times10^{-5}$, zero exceedances in $20{,}000$). At the pair level the same effect holds,
$\rho = 0.35$ ($n = 276$ pairs, permutation $p = 0.011$), rising monotonically across skill
quartiles from mean $r = 0.04$ in the weakest quartile to $r = 0.29$ in the strongest.

\paragraph{Is this just attenuation?} The obvious objection is statistical rather than
substantive. Correlation between two noisy measurements is attenuated toward zero, so weak
monitors may be unable to correlate highly with anything by construction. The confound is real in
our data, since skill and score spread on attack samples are themselves strongly related (Spearman
$0.78$), so we test it three ways. First, we recompute agreement as a \emph{rank} correlation,
which is invariant to score scale and spread and therefore cannot be produced by the wider-spread
mechanism, and the relationship is unchanged ($\rho = 0.818$, $p = 5\times10^{-5}$, versus $0.841$
for Pearson). Second, controlling for score spread explicitly reduces but does not remove it
(partial $\rho = 0.65$ for Pearson agreement, $0.58$ in rank space). Third, restricting to the 18
monitors above chance, which removes the degenerate floor, leaves $\rho = 0.68$ ($p = 0.003$).
Spread accounts for part of the effect, and most of it survives.

The interpretation is uncomfortable for ensemble monitoring. Two competent monitors agree because
they both find the real backdoors, while two weak monitors disagree because they are each wrong in
different places, so measured diversity in a weak pool is substantially noise. The next section
turns that reading into a measurement.

\subsection{The diversity metric fails because its two components cancel}\label{sec:mech}

Panels are built by minimising average pairwise correlation \citep[\S5.4]{koran2026ensemble}, so
the question that decides whether that is sound is a simple one. Does pairwise agreement predict
how much a panel gains over its best member? Across our 276 pairs, it essentially does not
(Spearman $+0.05$).

The reason is that agreement is a sum of two things that pull in opposite directions. Write a
monitor's standardised score on an attack item as $s_m(i) = \lambda_m\, d(i) + e_m(i)$, where
$d(i)$ is the item's intrinsic detectability (some backdoors are easy for everyone), $\lambda_m$
is how strongly the monitor tracks it, and $e_m$ is idiosyncratic error. Two monitors then agree
through a shared-signal term that grows with $\lambda_m\lambda_n$ and an error term
$\mathrm{corr}(e_m,e_n)$. We estimate $d$ by the leave-out mean score over all monitors except the
pair being measured, and split each pair's agreement into a signal part and an error part
(residual agreement after removing $d$).

Those two components predict panel gain with opposite sign, signal agreement at Spearman $-0.25$
and error agreement at $+0.26$, total agreement at $+0.05$. Summing them produces a metric whose
association with the thing it is used to optimise is near zero. We deliberately read no causal
story into either sign. Gain is measured against the panel's own best member, so both components
are entangled with pair skill and with pair redundancy, which we cannot separate here. Our claim
is the negative one and needs no such story. The composite is uninformative about gain because its
parts move in opposite directions, so a panel chosen by minimising it is not being steered toward
anything in particular. The cancellation is not a quirk of one evaluation. Across the two nested
monitor pools on task 1 and the six attacker slices on task 2, signal is negative and error
positive in 7 of 8, and in 7 of 8 the total sits closer to zero than either component
(Table~\ref{tab:cancel}).

\begin{table}[t]
\centering
\small
\caption{The cancellation, in every evaluation available. Cells are Spearman correlations with
the gain of a $k=2$ mean-aggregated panel over its own best member. The composite metric (total)
is the one panels are selected with, and it lies closer to zero than either of its parts in 7 of the
8 rows. \textsc{gpt-4o} is the exception, where the signal component is itself near zero.}
\label{tab:cancel}
\begin{tabular}{lccc@{\hskip 2em}lccc}
\toprule
Evaluation & Sig. & Err. & Tot. & Evaluation & Sig. & Err. & Tot. \\
\midrule
Task 1, 22 monitors & $-0.28$ & $+0.24$ & $+0.03$ & T2, GPT-4    & $-0.36$ & $+0.32$ & $+0.08$ \\
Task 1, 24 monitors & $-0.25$ & $+0.26$ & $+0.05$ & T2, GPT-4o   & $+0.01$ & $+0.33$ & $+0.27$ \\
T2, Claude-3.7      & $-0.28$ & $+0.23$ & $+0.02$ & T2, o1-mini  & $-0.27$ & $+0.38$ & $+0.13$ \\
T2, Gemini-2.0      & $-0.28$ & $+0.25$ & $+0.06$ & T2, o3-mini  & $-0.23$ & $+0.22$ & $+0.01$ \\
\bottomrule
\end{tabular}
\end{table}

The same decomposition explains \S3.1. \emph{Skill acts on signal:} the signal component rises
steeply with pair skill (Spearman $+0.53$, from $0.008$ to $0.219$ across skill quartiles) while
the error component is flat ($-0.006$, and non-monotone across quartiles). Competent monitors
agree because they both load on real detectability; weak monitors' apparent diversity is
idiosyncratic error.

The proxy invites a circularity objection. Because $d$ is built from the other monitors, so
``signal agreement rises with skill'' might restate ``strong monitors agree with a consensus they
dominate''. Estimating $d$ from a \emph{disjoint} half of the pool and measuring on the other
half, the signal--skill relationship holds in both directions ($+0.72$ and $+0.42$); the
error--skill relationship is unstable at $n=66$ per half and takes both signs, so we claim the
signal result and not a precise value for error. It does \emph{not} hold when $d$ comes from
below-median monitors only, where the relationship inverts. That is the expected failure mode,
since a consensus of near-chance monitors estimates noise rather than detectability. The
decomposition requires a consensus containing competent monitors, and we state that as a condition
on it.

\begin{figure}[t]
  \centering
  \includegraphics[width=0.88\textwidth]{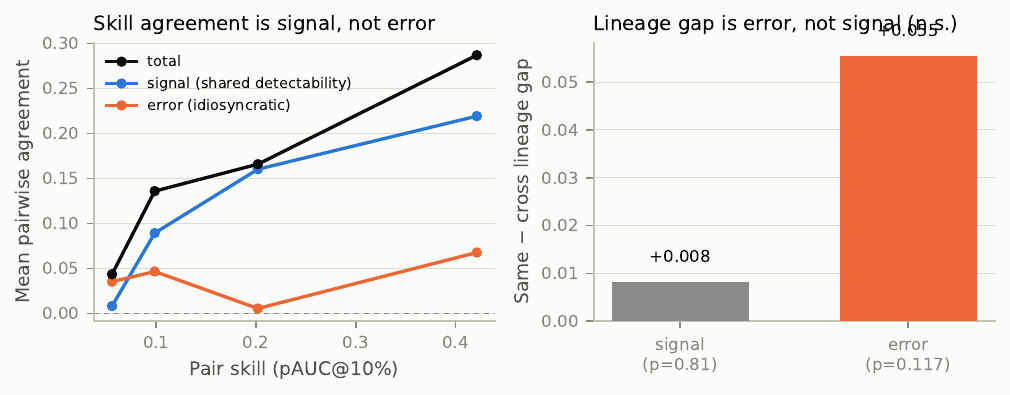}
  \caption{Left: agreement split into a shared-detectability signal component and an idiosyncratic
  error component, by pair-skill quartile. Skill raises signal; error stays flat. Right: the
  same-minus-cross lineage agreement gap, which is mostly error but is not significant in this
  pool. Skill acts on signal, lineage acts on error, and only signal is what detection needs.}
  \label{fig:mech}
\end{figure}

\subsection{Lineage moves neither detection nor, in a large enough pool, the metric}

If cross-lineage monitors were complementary in a way same-lineage monitors are not, they should
agree less, and panels built from them should detect more. Neither is measurable here.

\textbf{Detection.} Holding the anchor monitor fixed and varying only its partner, matched on
validation capability, a cross-lineage partner yields $-0.0084$ pAUC relative to a same-lineage
partner (pooled $n = 1938$ rows over six attacker evaluations). A monitor-level permutation of
lineage labels puts that comfortably inside the null ($p = 0.13$, null 95\% interval $[-0.012,
+0.010]$). This is the test to read, because the rows share anchor monitors, so the pair-level
bootstrap that an earlier version of this analysis reported, CI $[-0.011, -0.006]$, excluding
zero, is anticonservative, and its apparent significance is an artifact of treating dependent rows
as independent.

\textbf{The metric.} Pooling the six attacker evaluations, same-lineage pairs agree $0.064$ more
than cross-lineage pairs. All six per-attacker gaps are positive, spanning $+0.040$ to $+0.098$,
which looks like six confirmations and is not. The slices share monitors and share the honest
pool, so a monitor-level permutation that moves every slice at once is the dependence-respecting
test, and it returns $p = 0.18$. Decomposed, $87\%$ of the gap sits in the error component
($+0.055$, $p = 0.12$) against $+0.008$ in signal ($p = 0.81$). That is the direction the
mechanism predicts, at a magnitude this pool cannot resolve.

\textbf{Does it switch on among competent monitors?} The standing objection to a null pooled over
a mostly-weak pool is that lineage might matter where monitors actually work. Slicing to the
strongest monitors answers this with 3 rows and 2 anchors, which resolves nothing. Instead we ask
it of every row. The cross-minus-same difference among above-median anchors minus the same among
below-median anchors is $-0.021$ ($1026$ and $912$ rows; monitor-level permutation $p = 0.042$).
If anything, cross-lineage partners do slightly \emph{worse} among the more competent anchors.
This test was added after the pool grew and is therefore post-hoc, not pre-registered. We report
the sign for completeness, but a $p$ of $0.042$ on a single post-hoc interaction is weak evidence,
and the honest reading is that lineage does not help at either end.

\subsection{Correlation structure is a property of the pool}\label{sec:pools}

Twice in this project a result was invalidated by changing which monitors were in the pool, and
both times the earlier version was ours.

\textbf{The floor effect.} Our initial pool held monitors of 13B parameters and below, topping out
at pAUC 0.23 against the 0.4--0.8 range of \citet{koran2026ensemble}. On that pool we measured a
clean lineage null and were prepared to report it. Adding larger monitors changed the regime.
Gemma-4-26B reaches 0.781 and Qwen3.6-27B 0.803, roughly $4\times$ the old best member. We checked
both for harness artifacts rather than assuming. They hold on validation with its different
attacker mix (0.932 and 0.928 AUROC), code length alone scores chance (AUROC 0.47, $r \le 0.13$
with the score), 20/20 and 17/20 of their most-suspicious samples are true attacks, and neither is
degenerate. Mean pairwise agreement was $r \approx 0.18$ among the weak monitors; among the
competent ones it is $0.58$ across lineages.

\textbf{The lineage gap.} On the resulting 22-monitor pool, the same-minus-cross agreement gap was
$+0.104$ with a monitor-level permutation $p = 0.037$, and we wrote it up as a real effect. Adding
two monitors moved it to $+0.064$ at $p = 0.18$: Mistral-Small-24B at pAUC $0.407$ and OLMo-2-32B
at $0.166$, both in lineages that previously had only near-chance members. The mechanism is
visible in the two halves of the statistic. Same-lineage agreement barely moved ($0.247 \to
0.232$), while \emph{cross}-lineage agreement rose ($0.142 \to 0.168$). Adding one competent
monitor in a new lineage raises its agreement with every other lineage, because agreement tracks
skill (\S3.1). The apparent lineage effect was substantially a statement about which lineages
happened to contain the competent monitors.

The caution generalises beyond us. Any quantity depending on a monitor pool's correlation
structure, which means every ensemble-diversity result including the metric panels are selected
with, can move by a factor of three between a weak pool and a competent one, and can lose
significance when a single lineage gains a competent member. A null measured on weak monitors is
not evidence about deployed monitors, and an effect measured on a pool whose strong members
cluster in a few lineages is not evidence about lineage.

\subsection{Ensemble gain falls as pool skill rises}

\begin{figure}[t]
  \centering
  \includegraphics[width=0.52\textwidth]{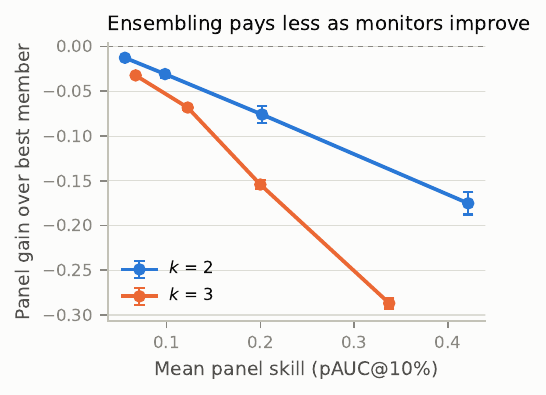}
  \caption{Panel gain over the best single member, by quartile of mean panel skill. Bars are
  standard errors. The advantage of ensembling shrinks as the monitors composing the panel get
  better.}
  \label{fig:gain}
\end{figure}

Sections 3.1--3.2 measure agreement, which is a proxy. Figure~\ref{fig:gain} measures the thing
itself: the gain of a mean-aggregated panel over its own best member, as a function of the panel's
mean skill. Gain falls monotonically across skill quartiles, from $-0.013$ to $-0.175$ at $k=2$ and
from $-0.032$ to $-0.287$ at $k=3$ (Spearman $-0.66$ and $-0.70$).

Two honest caveats. Gain is negative throughout, so in this pool an unselected panel never beats
its best member. That is consistent with \citet{anon2026diversity}, who find majority vote beats
the best member in 9.98\% of canonical size-3 subsets (1.27\% pooled over sizes 2--4). And part of
the decline is mechanical, since a stronger panel has a stronger best member, which is a higher
bar to clear. Neither caveat changes the direction, and the direction is the claim. Whatever
ensembling buys, it buys less as monitors improve.

\subsection{What still beats every ensemble we built}\label{sec:single}

Across our pool the single best configuration is not an ensemble, and this holds in two separate
regimes. Within the six prompt variants on a fixed base (Qwen3-8B), the best single variant
reaches pAUC 0.228 against 0.226 for the best prompt-only panel. Over the full 24-monitor pool,
the best single monitor reaches 0.803 against 0.808 for the best two-model panel and 0.793 for the
best mixed prompt-and-model panel. Every one of these is a maximum over an exhaustive test-split
search with no held-out selection, so a 0.005 margin is selection noise and we claim only the
qualitative statement that no panel we built beats a single well-chosen monitor by more than
selection noise. Section~\ref{sec:sel2} makes the same point under an honest out-of-sample
protocol, where the single best monitor wins outright.

One clarification matters for comparability. Our within-model panels show a \emph{negative} mean
gain ($-0.032$ at $k=2$), which might look like a failure to reproduce \citet{koran2026ensemble}'s
positive ensemble effect. It is not: we measure gain over the panel's own best member, a strictly
harder baseline than theirs, which is a single baseline monitor. Measured their way, against our
baseline prompt variant at pAUC $0.149$, the best prompt-only panel at $0.226$ is a $52\%$
relative gain, and their effect reproduces. The two framings disagree only about the reference
point, and we report the stricter one throughout.

Separately, prompt choice on fixed weights moved pAUC by $3.3\times$ (0.069 to 0.228), comparable
to the $5.4\times$ spread across the 14 monitors that clear chance on validation but fall below
the competence threshold (0.036 to 0.193): down there, how a monitor is prompted is worth about as
much as which one is chosen.

\subsection{No correlation-weighted selection beats the best single monitor}\label{sec:sel2}

The sharper test is whether the diversity metric helps \emph{select} a panel out of sample. We
select panels on the single-attacker data by the objective $\overline{\text{capability}} -
\lambda\,\overline{\text{agreement}}$, which is skill-only at $\lambda=0$ and pure
diversity-seeking as $\lambda\to\infty$, then evaluate the chosen panel on the six-attacker set it
never saw. No positive $\lambda$ reliably helps. At $k=3$ the out-of-sample pAUC is flat at
$0.723$ from $\lambda=0$ through $\lambda=1$, then falls to $0.440$ at $\lambda=4$ and $0.068$ at
pure diversity; at $k=4$ it is flat at $0.676$ through $\lambda=0.5$ and falls thereafter. Against
this, the single best monitor chosen by validation capability scores $0.733$ out of sample, higher
than every selected panel at either $k$. Selecting for decorrelation does not survive contact with
a new task. Skill selection is both simpler and, out of sample, at least as good.

\subsection{Everything replicates across attackers}

The findings above are not artifacts of one attacker. Agreement rises with skill in five of six
attacker evaluations ($\rho$ from $+0.39$ to $+0.85$; the exception is GPT-4, our smallest slice
at $n = 57$, where $p = 0.06$). Panel gain falls with pool skill in six of six ($\rho$ from
$-0.58$ to $-0.67$). The signal/error cancellation of \S\ref{sec:mech} holds in five of six
(Table~\ref{tab:cancel}). And monitor skill itself is nearly invariant to who wrote the backdoors.
A monitor's skill on the single-attacker evaluation predicts its mean skill across the
six-attacker evaluation at $\rho = 0.95$ ($p < 10^{-5}$, $n = 24$). Whatever these monitors are
detecting, it is not specific to one attacker's style.

\section{Implications}

Ensemble monitoring is motivated by the intuition that independent signals cover each other's
blind spots. Our measurements suggest that intuition is strongest exactly where it matters least.
The pool where diversity is abundant is the pool of weak monitors, whose absolute detection is
poor. As monitors become competent enough to deploy, they converge, and the diversity available to
exploit falls away. This does not contradict \citet{koran2026ensemble}, whose ensembles do beat
their single monitors, but it bounds the extrapolation. the ensemble advantage should not be
assumed to persist as monitors improve, and it should be re-measured at the capability of the
monitor actually being deployed.

The sharper consequence is about the metric. Average pairwise correlation is the instrument used
to build diverse ensembles, and \S\ref{sec:mech} shows it sums two terms whose associations with
gain point in opposite directions, which is why it barely predicts the gain it is used to
optimise. The obvious repair is to select on the error component alone, and we tested that rather
than recommending it. Substituted into the out-of-sample scan of \S\ref{sec:sel2} it does
\emph{not} help at either panel size, matching skill-only selection at small weights and falling
faster at large ones ($0.646$ against $0.723$ at $k=3$, $\lambda=1$). Neither beats the single
best monitor. Decomposing the metric explains its failure. It does not repair it.

What survives as advice is smaller and better supported. Spend the first unit of effort on
prompting a competent monitor well, not on adding a second lineage. Prompt choice moved detection
by more than $3\times$ on fixed weights, while adding a lineage bought neither decorrelation nor
detection. And before trusting any published diversity number, including ours, ask which monitors
were in the pool.

\section{Limitations}\label{sec:limits}

This is a single task family: backdoored competitive-programming solutions, from one dataset,
scored one way (\textsc{yes}/\textsc{no} decision log-odds). Varying the attacker across six
models addresses who wrote the sabotage, not what kind of sabotage it is. A genuinely different
task, such as sabotage in long-horizon agent trajectories, remains untested and is the experiment
we would run next.

Our lineage results are nulls, and a null is a statement about power. With one to six monitors per
lineage we can exclude a cross-lineage detection benefit of roughly the size of one attack item.
At about 128 attacks per attacker slice, one item moves normalised pAUC@10\% by around $0.008$. We
cannot exclude a smaller one. The monitor-level permutation null for the matched-partner statistic
spans $[-0.012, +0.010]$ pAUC, and effects inside that band are invisible to us. A pool with many
competent monitors per lineage would settle it, and a 65\,GiB single-node model-loading ceiling
bounds how large those members can be. We think the more useful reading of \S\ref{sec:pools} is
not ``lineage does nothing'' but ``this statistic is unstable in pools of this size'', which is
also a warning about the positive lineage results in the literature.

The signal/error decomposition rests on a leave-out consensus as the estimate of item
detectability, and \S\ref{sec:mech} shows it fails when that consensus contains no competent
monitors. It is a measurement device with a stated operating range, not a general-purpose
estimator.

Finally, all monitors here are open-weight models of at most 32B. Whether the agreement--skill
relationship continues into frontier-scale monitors is untested, and it is the regime that matters
most for deployment. Our claim is that ensemble headroom shrinks as monitors improve, and the
models actually being deployed as trusted monitors sit above the top of our range.

\section{Related work}

\citet{koran2026ensemble} is the direct antecedent and the source of our data, metric, prompt and
scoring design. Our contribution is to vary the axes their single-base-model design holds fixed.
\citet{pinto2026calibration} make untrusted model family the controlled axis for single monitors
and find a real own-family interaction, consistent with ours. They vary the family of the
\emph{monitored} model with a trained monitor, whereas we vary the lineage of the \emph{monitors}
themselves. \citet{anon2026diversity} audits diversity metrics for majority-vote ensembles under
capability control and finds strict diversity nearly collinear with member capability. Our \S3.1
is the weak-challenge analogue in a monitoring setting, and \S\ref{sec:mech} goes past
collinearity to say which part of the metric carries the confound. Three concurrent results in the
\emph{voting} setting converge with ours. \citet{cohen2026dfpe} filter for competence before
selecting for diversity, the prune-then-diversify discipline our selection scan
(\S\ref{sec:single}) finds necessary but not sufficient out of sample. \citet{chen2026cofailure}
show across 67 frontier models that combining rarely beats the single best model absent a
query-level routing signal, which is convergent evidence for our selection result at far larger
scale. Closest to our mechanism, \citet{ali2026diversity} find raw correctness correlation has
almost no predictive power ($R^2 \le 0.09$) while an accuracy-adjusted variant reaches $R^2 =
0.67$. That is the same insight, that raw agreement confounds capability, reached independently in
the voting setting. Our contribution relative to that line is to \emph{decompose} agreement rather
than correct a scalar, and to show the parts predict gain with opposite sign. Trusted monitoring
follows \citet{greenblatt2024control}.

All monitors are open-weight, all data is public, and every number here is written by the analysis
harness to a results file. \cpanartifact



\begin{thebibliography}{9}
\bibitem[Ali(2026)]{ali2026diversity}
J.~Ali.
\newblock Quantifying diversity of thought: A predictive law of weighted LLM ensemble lift.
\newblock \emph{\arxivref{2607.17384}}, 2026.

\bibitem[Chen(2026)]{chen2026cofailure}
J.~Chen.
\newblock When does combining language models help? A co-failure ceiling on routing,
  voting, and mixture-of-agents across 67 frontier models.
\newblock \emph{\arxivref{2606.27288}}, 2026.

\bibitem[Cohen et al.(2026)]{cohen2026dfpe}
S.~Cohen, N.~Cohen Inger, N.~Goldshlager, B.~Shapira, and L.~Rokach.
\newblock DFPE: A diverse fingerprint ensemble for enhancing LLM performance.
\newblock In \emph{Findings of the ACL: EACL 2026}, pages 5326--5336, 2026.
  \arxivref{2501.17479}.

\bibitem[Greenblatt et al.(2024)]{greenblatt2024control}
R.~Greenblatt, B.~Shlegeris, K.~Sachan, and F.~Roger.
\newblock AI control: Improving safety despite intentional subversion.
\newblock \emph{ICML}, 2024. \arxivref{2312.06942}.

\bibitem[Jha(2026)]{jha2026experts}
A.~Jha.
\newblock Half the experts, all the code: One-shot domain pruning of mixture-of-experts LLMs for
  coding.
\newblock \emph{\arxivref{2607.16721}}, 2026.

\bibitem[Kim(2026)]{anon2026diversity}
D.~Kim.
\newblock Are diversity metrics measuring diversity? A capability-controlled audit of
majority-vote gain in LLM ensembles.
\newblock \emph{\arxivref{2607.20768}}, 2026.

\bibitem[Koran et al.(2026)]{koran2026ensemble}
E.~Koran, Y.~Yun, S.~Tetef, B.~Arnav, and P.~Bernabeu-Perez.
\newblock Ensemble monitoring for AI control: Diverse signals outweigh more compute.
\newblock \emph{\arxivref{2605.15377}}, 2026.

\bibitem[Pinto(2026)]{pinto2026calibration}
L.~Pinto.
\newblock Calibration-family overfit: Why trusted sabotage monitors don't transfer across lineages.
\newblock \emph{\arxivref{2607.06596}}, 2026.
\end{thebibliography}
\end{document}